\documentclass[]{pasj01}

\begin{document} 
\Received{}
\Accepted{}

\title{Mid-infrared Multi-wavelength Imaging of Ophiuchus\,IRS\,48\,Transitional Disk}

\author{Mitsuhiko \textsc{Honda}\altaffilmark{1}%
\thanks{Based on data collected at Subaru Telescope, which is operated by the National Astronomical Observatory of Japan }}
\altaffiltext{1}{Department of Physics, Kurume University School of Medicine, 
67 Asahi-machi, Kurume, Fukuoka, 830-0011, Japan}
\email{hondamt1977@gmail.com}

\author{Kazushi \textsc{Okada},\altaffilmark{2}}
\altaffiltext{2}{Institute of Astronomy, Graduate School of Science, The University of Tokyo,
Osawa 2-21-1, Mitaka, Tokyo 181-0015, Japan}

\author{Takashi \textsc{Miyata}\altaffilmark{2}}

\author{Gijs D. \textsc{Mulders}\altaffilmark{3}}
\altaffiltext{3}{Lunar and Planetary Laboratory, The University of Arizona, Tucson, AZ 85721, USA}

\author{Jeremy R. \textsc{Swearingen}\altaffilmark{4}}
\altaffiltext{4}{Department of Physics, University of Cincinnati, 400 Geology/Physics Building, P.O. Box 210011, Cincinnati, OH 45221-0377, USA}

\author{Takashi \textsc{Kamizuka}\altaffilmark{2}}

\author{Ryou \textsc{Ohsawa}\altaffilmark{2}}

\author{Takuya \textsc{Fujiyoshi}\altaffilmark{5}}
\altaffiltext{5}{Subaru Telescope, National Astronomical Observatory of
Japan, 650 North A'ohoku Place, Hilo, Hawaii 96720, U.S.A.}

\author{Hideaki \textsc{Fujiwara}\altaffilmark{5}}

\author{Mizuho \textsc{Uchiyama}\altaffilmark{6}}
\altaffiltext{6}{National Astronomical Observatory of Japan, 2-21-1 Osawa, Mitaka, Tokyo 181-8588, Japan}

\author{Takuya \textsc{Yamashita}\altaffilmark{6}}

\author{Takashi \textsc{Onaka}\altaffilmark{7}}
\altaffiltext{7}{Department of Astronomy, Graduate School of Science, The University of Tokyo, Bunkyo-ku, Tokyo 113-0033, Japan}





\maketitle

\begin{abstract}
\noindent
Transitional disks around the Herbig Ae/Be stars are fascinating targets in the contexts of 
disk evolution and also planet formation. Oph IRS 48 is one of such Herbig Ae stars, 
which shows an inner dust cavity and azimuthally lopsided large dust distribution. 
We present new images of Oph IRS 48 at eight mid-infrared (MIR) wavelengths from 8.59 to 24.6\,$\mu {\rm m}$ 
taken with the COMICS mounted on the 8.2\,m Subaru Telescope. 
The N-band (7 to 13\,$\mu {\rm m}$) images show that the flux distribution is centrally peaked with a slight spatial extent, 
while the Q-band (17 to 25\,$\mu {\rm m}$) images show asymmetric double peaks (east and west). Using 18.8 and 24.6\,$\mu$m images, we derived the dust temperature at both east and west peaks to be 135$\pm$22 K.
Thus, the asymmetry may not be attributed to a difference in the temperature.
Comparing our results with previous modeling works, 
we conclude that the inner disk is aligned to the outer disk.
A shadow casted by the optically thick inner disk has a great influence on the morphology
of MIR thermal emission from the outer disk.
\end{abstract}

\section{Introduction} \label{sec:intro}

Circumstellar disks around Herbig Ae/Be stars are one of the most interesting targets 
to study the environments of planetary birthplaces. The Herbig Ae/Be stars are the 
intermediate-mass counterparts ($> \sim 2 M_\odot$) of the T Tauri stars, with spectral types
A and B \citep{her60a,wat98a}. Circumstellar disks around the Herbig Ae/Be stars have wide diversity among objects.
Spatially unresolved spectral energy distribution (SED) modeling on these disks indicated that
some objects have a disk with significantly depleted inner regions 
(i.e., pre-transitional or transitional disks; \cite{str89a,skr90a,esp07a}). 
Actually, recent high spatial resolution observations reveal more complex disk structures 
such as spirals, warps, and azimuthally lopsided structures as well as gaps and dips (for a review, see \cite{esp14a}).
Such an inner gap or a dip in a protoplanetary disk can be interpreted by a variety of theoretical 
explanations such as grain growth \citep{tan05a,dul05a,bir12a}, photo-evaporation \citep{cla01a,ale06a}, 
photophoresis \citep{kra07a}, magneto-rotational instability \citep{chi07a}, 
dynamical interaction with stellar or substellar companions \citep{lin79a,lin79b,art94a}. 
In addition to these possibilities, forming planets can also be a convincing disk clearing 
mechanism \citep{lin86a,bry99a,zhu11a,ric06a,zhu12a,pin12a}. 
The inner gaps and/or holes of (pre-)transitional disks are attractive objects not only 
for understanding the evolution of disks but also for revealing the formation history of planets. Some of 
the cleaning mechanisms mentioned above have a direct link to the forming planets, and the others 
potentially influence on the forming process of young planets.

A dozen or two dozen of (pre-)transitional disks have been extensively studied.
These studies are mainly based on near-infrared (NIR) polarized images taken 
with AO system and high resolution radio maps. 
Although they have a big advantage to resolve the detailed structure of the disks, 
{\bf NIR and radio emission do not properly constrain the warm dust in the inner region.}
On the contrary, MIR observations detect the thermal emissions from the 
inner regions, and, by using a ground-based large telescope, can achieve high spatial resolution to resolve 
the inner regions of nearby Herbig Ae/Be stars. They should play a key role in studying environments of 
the inner region, where planets have formed or will form. So far there are several Herbig Ae/Be 
stars whose disk structures have been well-resolved in the MIR wavelengths, 
such as HD 142527 \citep{fuj06a}, HD 97048 \citep{lag06a},
and Oph IRS 48 \citep{gee07a}. This paper focuses mainly on the Oph IRS 48 transitional disk.

Oph\,IRS\,48 (WLY\,2-48; \cite{wil89a}) is an A0 star 
\citep{bro12a,fol15a} in the $\rho$\,Oph region, at a distance of 120 pc \citep{loi08a}. 
The mass and age of the star are estimated to be 2.0 to 2.5 $M_\odot$ and 5 to 15 Myr old, respectively 
\citep{bro12a,fol15a}. 
It has attracted a great deal of attention mainly because of its highly asymmetric 
sub-millimeter dust continuum observed by ALMA \citep{mar13a}. 
Such an azimuthally lopsided flux distribution may be a consequence of a particle trap 
in the local gas pressure maxima. The particle trap can be a key to overcoming obstacles 
to the complete planet formation theory, including the radial drift of particles (e.g., \cite{wei77a}).

Resolved observations towards Oph IRS 48 were firstly made with the VISIR instrument 
on the Very Large Telescope (VLT) at 8.6, 9.0, 11.3, 11.9 and 18.7\,$\mu$m \citep{gee07a}. 
In the {\it N}-band (7.5 to 13\,$\mu$m) images, the emission is strongly peaked at the central star and thought to originate 
mostly from polycyclic aromatic hydrocarbons (PAHs) and/or very small carbonaceous grains (VSGs). 
On the other hand, it shows a ring-like structure in the 18.7\,$\mu$m image with a radius of approximate 60 AU and a cavity in the center, suggesting that the\,$\mu$m-sized dust is almost depleted in the central ($\lesssim 60$ AU) region \citep{maa13a}. 
Further spatially resolved observations were made by \citet{bro12a,bro12b}, \citet{bru14a}, \citet{fol15a}, \citet{mar15}, and \citet{mar16}. 
These observations covered a wide wavelength range from NIR to centimeter wavelengths. 
Utilizing these observations, \citet{maa13a}, \citet{bru14a}, and \citet{fol15a} 
presented SED models of Oph IRS 48, but interpretations of the observed disk geometry are somewhat different. 
\citet{maa13a} proposed an optically thin halo as an inner structure in the cavity, 
while \citet{bru14a} and \citet{fol15a} suggested an optically thick disk.

This paper present new multi-wavelength images of Oph IRS 48 at MIR wavelengths.
We compare our results with the previous observations 
and the modeling works, and discuss the observed geometry, especially the inner structure of the disk, 
which strongly influences MIR appearances. We also compare our results with HD 142527, 
another well-studied Herbig Ae star, to develop a more uniform understanding of the transitional disks.

\section{Observations and Data Reduction}  \label{sec:obs}

\subsection{Observations}

Oph\,IRS\,48 was observed with the COMICS \citep{kat00a,oka03a,sak03a} 
mounted on the 8.2 m Subaru Telescope on 2015 April 26 and 27. The pixel scale of the COMICS 
is 0.13 arcsec pixel$^{-1}$. The observations were made in five {\it N}-band filters and three {\it Q}-band filters. 
Detailed properties of these filters are listed in Table \ref{filters}. Standard stars were also observed immediately 
before and after the observations of the science target at all the filters except for 8.8 and 12.4\,$\mu$m. 
In these filters, the standard stars were observed only before the target observations. The stars were 
selected from the standard star catalogue provided by \citet{coh99a}, considering elevations of the 
target and standard stars at the time. Table \ref{obssummary} displays a log of the observations. To cancel out the variations 
of background radiation, a chopping observing scheme is applied at a frequency of 0.45 Hz with an 8 arcsec 
throw. Because the residual pattern after chopping subtraction is negligible compared to the observing target 
for the Subaru/COMICS when the target is sufficiently bright, a nodding scheme was not adopted for both 
observations of the science target and the standard stars.

\begin{table}
  \tbl{Properties of filters}{%
  \begin{tabular}{rrc}
\hline
\multicolumn{2}{c}{Filter Wavelength ($\mu {\rm m}$)} & Atmospheric Window \\
Center & Width & \\
\hline
8.59 & 0.42 & $N$ band \\
8.8 & 0.8 & $N$ band \\
10.46 & 0.16 & $N$ band \\
12.4 & 1.2 & $N$ band \\
12.81 & 0.21 & $N$ band \\
17.7 & 0.9 & $Q$ band \\
18.8 & 0.9 & $Q$ band \\
24.6 & 0.8 & $Q$ band \\ \hline
  \end{tabular}}\label{filters}
\end{table}

\begin{table*}
  \tbl{Summary of the observations}{%
  \begin{tabular}{llllrr}
\hline
Date & Filter & \multicolumn{2}{c}{Object} & Integration time & Air-mass \\ 
  & $\mu {\rm m}$ & Name & Type & [sec.] & \\ 
\hline
2015 Apr 27 & 8.59 & ${\nu}$\,Lib & standard & 60 & 1.56-1.57 \\
 &  & Oph\,IRS\,48 & target & 803 & 1.48-1.64 \\
 &  & 9\,Her & standard & 60 & 1.43-1.44 \\
2015 Apr 26 & 8.8 & ${\delta}$\,Oph & standard & 33 & 1.09-1.13 \\
 &  & Oph\,IRS\,48 & target & 33 & 1.40 \\
2015 Apr 27 & 10.46 & 9\,Her & standard & 61 & 1.46-1.47 \\
 &  & Oph\,IRS\,48 & target & 550 & 1.73-2.09 \\
 &  & 9\,Her & standard & 183 & 1.83-1.92 \\
2015 Apr 26 & 12.4 & ${\delta}$\,Oph & standard & 35 & 1.09-1.14 \\
 &  & Oph\,IRS\,48 & target & 35 & 1.40 \\
2015 Apr 27 & 12.81 & 42 Lib & standard & 33 & 1.48-1.49 \\
 &  & Oph\,IRS\,48 & target & 662 & 1.43-1.58 \\
 &  & ${\delta}$ Oph & standard & 65 & 1.38 \\
2015 Apr 26 & 17.7 & ${\delta}$\,Oph & standard  & 180 & 1.25-1.27 \\
 &  & Oph\,IRS\,48 & target & 1261 & 1.63-2.04 \\
 &  & ${\delta}$ Oph & standard & 300 & 1.63-1.72 \\
2015 Apr 27 & 18.8 & ${\delta}$\,Oph & standard & 201 & 1.09-1.10 \\
 &  & Oph\,IRS\,48 & target & 1406 & 1.40-1.42 \\
 &  & ${\delta}$ Oph & standard & 201 & 1.13-1.14 \\
2015 Apr 26 & 24.6 & ${\delta}$\,Oph & standard  & 182 & 1.09-1.11 \\
 &  & Oph\,IRS\,48 & target & 1214 & 1.40-1.48 \\
 &  & ${\delta}$\,Oph & standard & 182 & 1.19-1.21 \\ \hline
  \end{tabular}}\label{obssummary}
  \begin{tabnote}
The observations are arranged in the order of the filter center wavelength and, for each wavelength, listed in time series.
  \end{tabnote}
\end{table*}

\subsection{Data Reduction}
The reduction of the obtained data was performed in a standard way of the COMICS data reduction with 
using standard IRAF tasks and our own tools. We also applied two additional procedures to reduce artificial 
pattern which were found in the images after the chopping subtraction. One is channel noise subtraction. 
The detector used in the COMICS has 16 channel outputs, and 16-pixel data are read out at the same time. 
If the extra noise intruded in the read-out system, it causes repetitive pattern noise in the images. To 
reduce this pattern noise, we subtracted the median value over the 16 pixels from each pixel. We note 
that this procedure did not cause any effect on the stellar images because they appeared in a 
small portion of the detector.

\begin{figure}
\begin{center}
 \includegraphics[width=80mm]{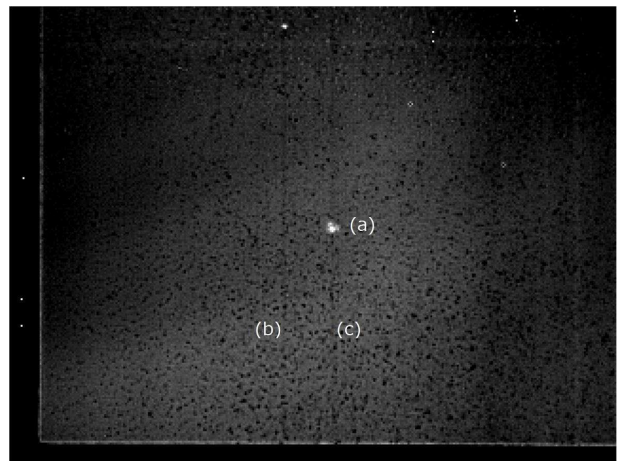}
\end{center}
 \caption{An 8.8-$\mu {\rm m}$ raw image frame. A bright source located at position (a) is a standard $\delta$ Oph. 
Two kinds of artificial vertical dark patterns located at columns (b) and (c) affect results. The former appears in initial 
several frames of each image file at the same position of the detector. If observing source is bright, 
the latter appears at the columns where the source is located.}
\label{figure1}
\end{figure}

Another procedure is vertical noise subtraction. Figure \ref{figure1} shows an example of the chopping subtracted image. 
Two vertical lines can be seen in this image. One is grew down from a hot pixel seen in the upper side of 
the image (labeled as b), and the other is associated with the bright stellar image (labeled as c). Both lines 
were cancelled out by subtracting the median of each pixel's column in the sky region from each pixel \citep{sak03a}.

After these procedures, a shift-and-adding scheme was applied to improve the blurring caused by tracking 
errors and fluctuating atmosphere. The data consisted of 0.98 sec source integration frames. The source 
position can be recognized in each frame since the science target and the standard stars are bright enough 
for all the wavelengths. We measured centroid position of the source in each frame, shifted them so as to align 
the centroid position, and summed up the frame into the final image. SExtractor \citep{ber96a} 
was used to detect centroids of sources.

\subsection{Flux Calibrations}
Flux calibration was performed with photometry of the standard stars for each filter. 
The intrinsic fluxes of the standard stars within the filter wavelengths were evaluated from 
Cohen's spectral templates \citep{coh99a}. In addition, air-mass corrections were made 
for the {\it Q}-band observations because the {\it Q}-band atmospheric transmittance is expected to be 
sensitive to the air-mass $X$. Assuming that the observed flux is reduced according to a function
$A \exp{(-B X)}$, the constants $A$ and $B$ at each wavelength can be derived from fitting 
two observation points. We used the photometric results of the standard stars observed before 
and after the target observation for the fitting. The correction was done for the science target observations by extrapolating the function.
As a result, this correction increased the flux 
densities at 17.7, 18.8, and 24.6\,$\mu$m by 2-6, 9-11, and 9-15 percent, respectively.

Uncertainty of the calibrated flux of each filter was estimated as follows. In the filters except 
the 24.6\,$\mu$m filter, signal-to-noise ratios (SNRs) of the standard stars were achieved more than 
100. Therefore, uncertainties from the photometry were 1\% or lower. On the other hand, 
the intrinsic fluxes calculated from the Cohen templates have 2 to 6\% uncertainties in this wavelength 
range. Consequently, the uncertainties of the flux were estimated as 2-6\% in total. In the 24.6\,$\mu$m filter, 
the SNRs were 50 and 20 for the data taken before and after the target observation, respectively. 
It was almost comparable to the uncertainty of the intrinsic flux (3\% at this wavelength). Therefore, 
we concluded that the flux uncertainty of the 24.6\,$\mu$m filter was 6\% in total. Note that the estimated 
uncertainties did not include the error from the air-mass calibration described above.

\subsection{Point Spread Functions}
Every standard star should be observed as a point source because the angular diameters 
of the standard stars \citep{coh99a} are at least one order of magnitude smaller than 
the pixel scale. To evaluate the spatial resolution achieved during the observations, a point 
spread function (PSF) is generated by calculating the average of the normalized standard star 
images observed before and after the target in every filter except the 8.8 and 12.4\,$\mu$m filters. 
In these filters, the standard stars were observed just before the target.
The FWHMs of the PSFs are 0.25, 0.27, 0.30, 0.32, 0.36, 0.45, 0.48, and 0.63 arcsec, at 8.59, 8.8, 10.46, 12.4, 12.81, 17.7, 18.8, and 24.6\,$\mu$m respectively.

\subsection{Position Alignment among Different Filters}
The position shifts in the image can occur among different filters due to the difference of optical and mechanical configuration.
Since the science target is an extended object and there are no point sources detected around it especially in Q-band, it is not easy to align the target positions among the images in different filters. 
To obtain accurate position shift information between filters, we sequentially observed the standard star in a short period of time in all the filters. 
We measured the position shifts of the standard star images, and applied it to the images 
of the science target assuming that the peak of the 8.59\,$\mu$m science image is the stellar position.
In the resulting image of each filter, the science target lies in its 
proper relative position to each other. 
The accuracy of the position shift is less than 0.2 arcsec in all the filters except 24.6\,$\mu$m, where it is nearly 0.3 arcsec. The aligned images of IRS\,48 are reasonably consistent 
with each other; they show a strong peak at the same position within the accuracy range 
in the {\it N}-band (Figure \ref{IRS48images}). In the 10.46\,$\mu$m filter, the sequential measurements were not conducted, 
and the position was aligned so as to match the peak position of IRS\,48.

\section{Results}

\subsection{{\it N}-band imaging}

Figure \ref{IRS48images} displays the images of Oph\,IRS\,48 obtained by the COMICS. 
Images of the standard stars are also displayed in the right panels as references of the PSFs.
 All of the five {\it N}-band (8.59 to 12.81\,$\mu$m) images show single-peak distributions, 
and their peak positions are consistent with each other. 
All images are clearly extended compared to the PSFs. 
At shorter wavelengths (8.59 and 8.8\,$\mu$m) the images have elliptical shapes whose centers 
are located at the peak position. On the other hand, the images at longer wavelengths 
(10.46, 12.4, and 12.81\,$\mu$m) are extended towards the north and shaped as an upside-down 
triangle. These appearances almost agree with the MIR images previously reported by 
\citet{gee07a}. 

\begin{figure*}
\begin{center}
 \includegraphics[width=160mm]{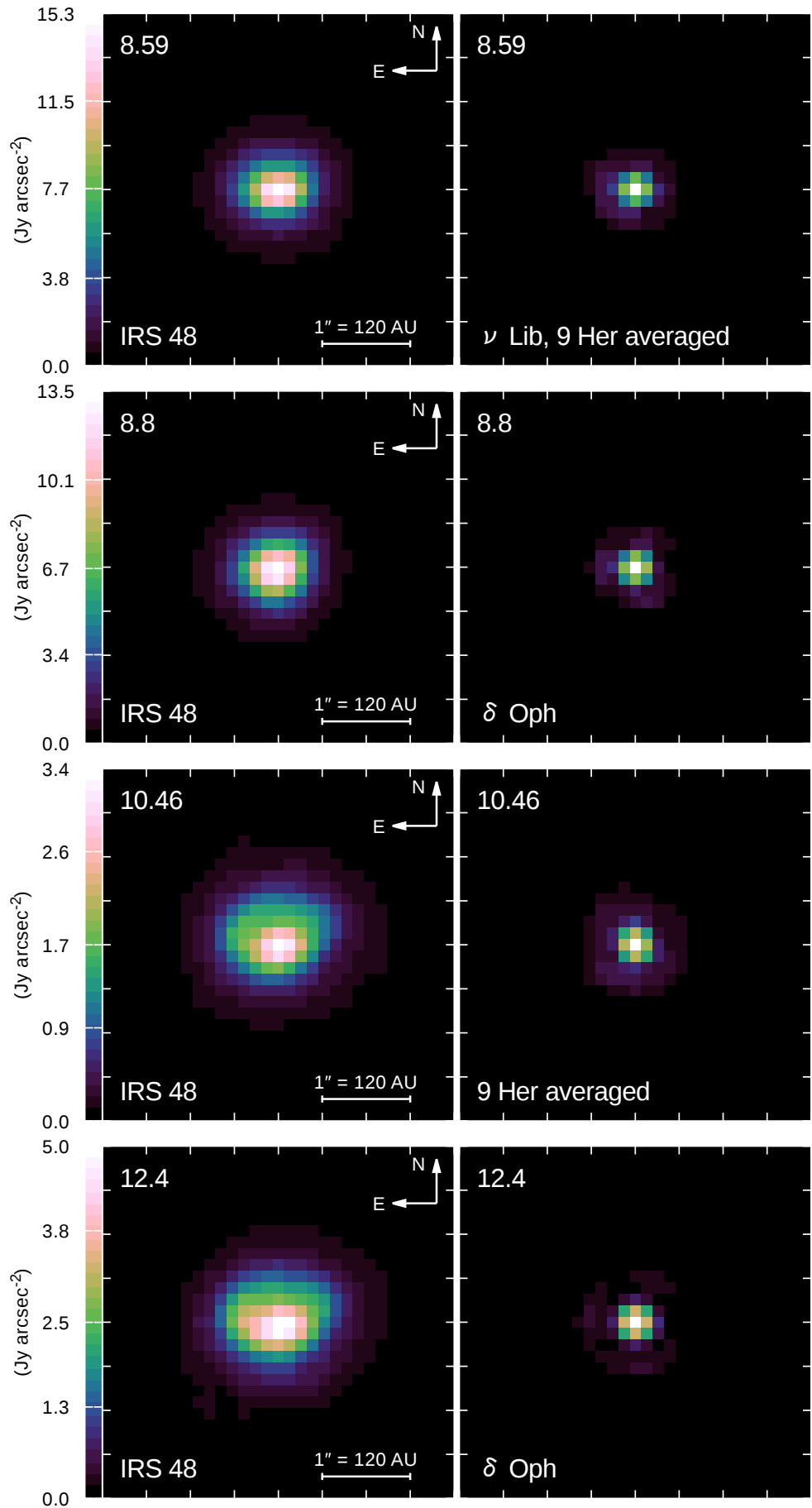}
\end{center}
 \caption{Subaru/COMICS images of Oph\,IRS\,48 and PSFs. The number in the upper-left 
corner of each panel is the filter center wavelength in $\mu$m. Left: the observed intensity of 
IRS\,48. We assume that the peak of the 8.59\,$\mu$m image is the star position, which is located 
at the center of each image. Right: the PSF for each wavelength. Except for at 8.8 and 12.4\,$\mu$m, 
every PSF is the average of the normalized standard star images taken before and after the target.}
\label{IRS48images}
\end{figure*}

\setcounter{figure}{1}
\begin{figure*}
\begin{center}
 \includegraphics[width=160mm]{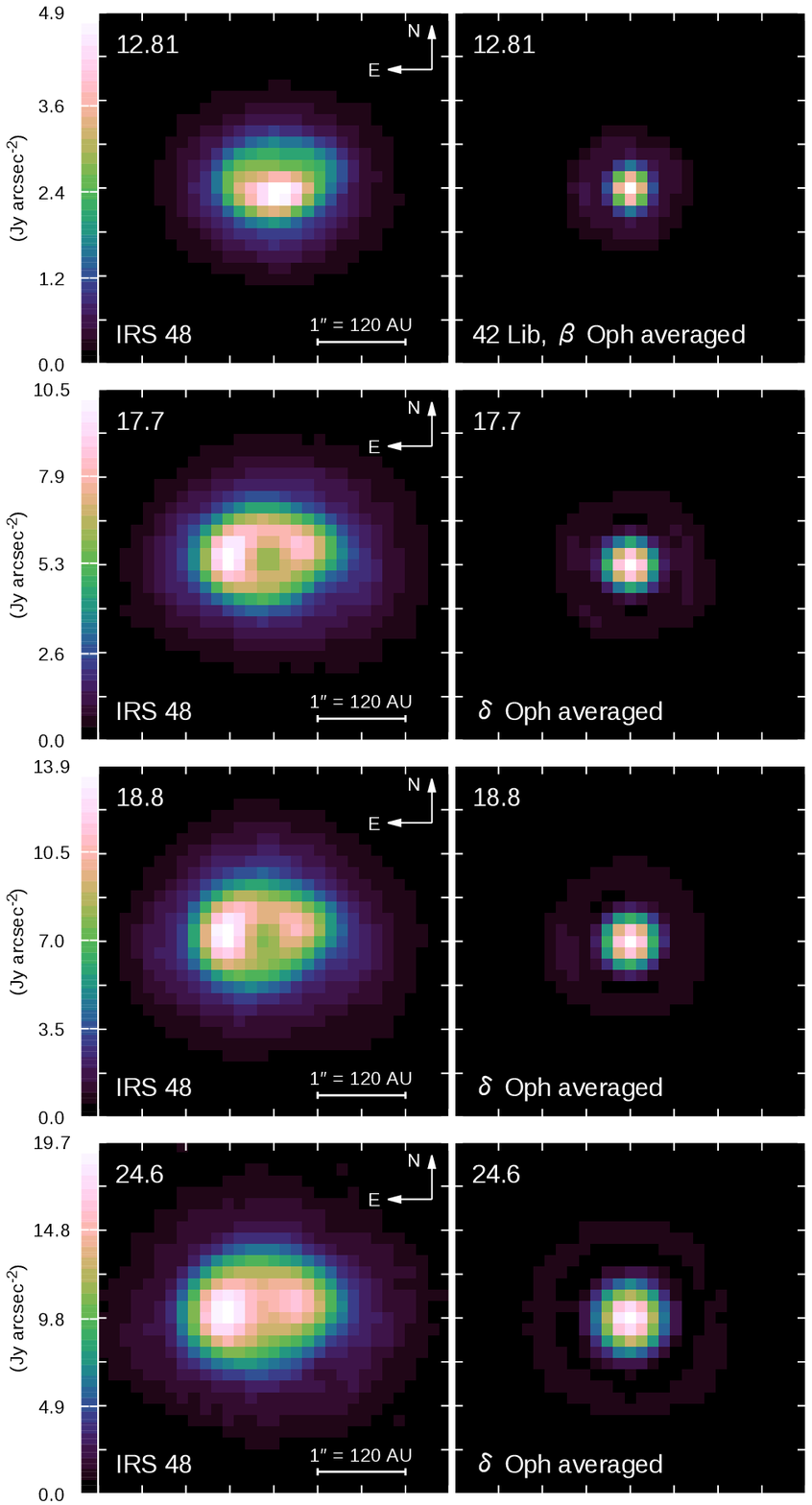}
\end{center}
 \caption{(Continued)}
\end{figure*}

\begin{figure*}
\begin{center}
 \includegraphics[width=160mm]{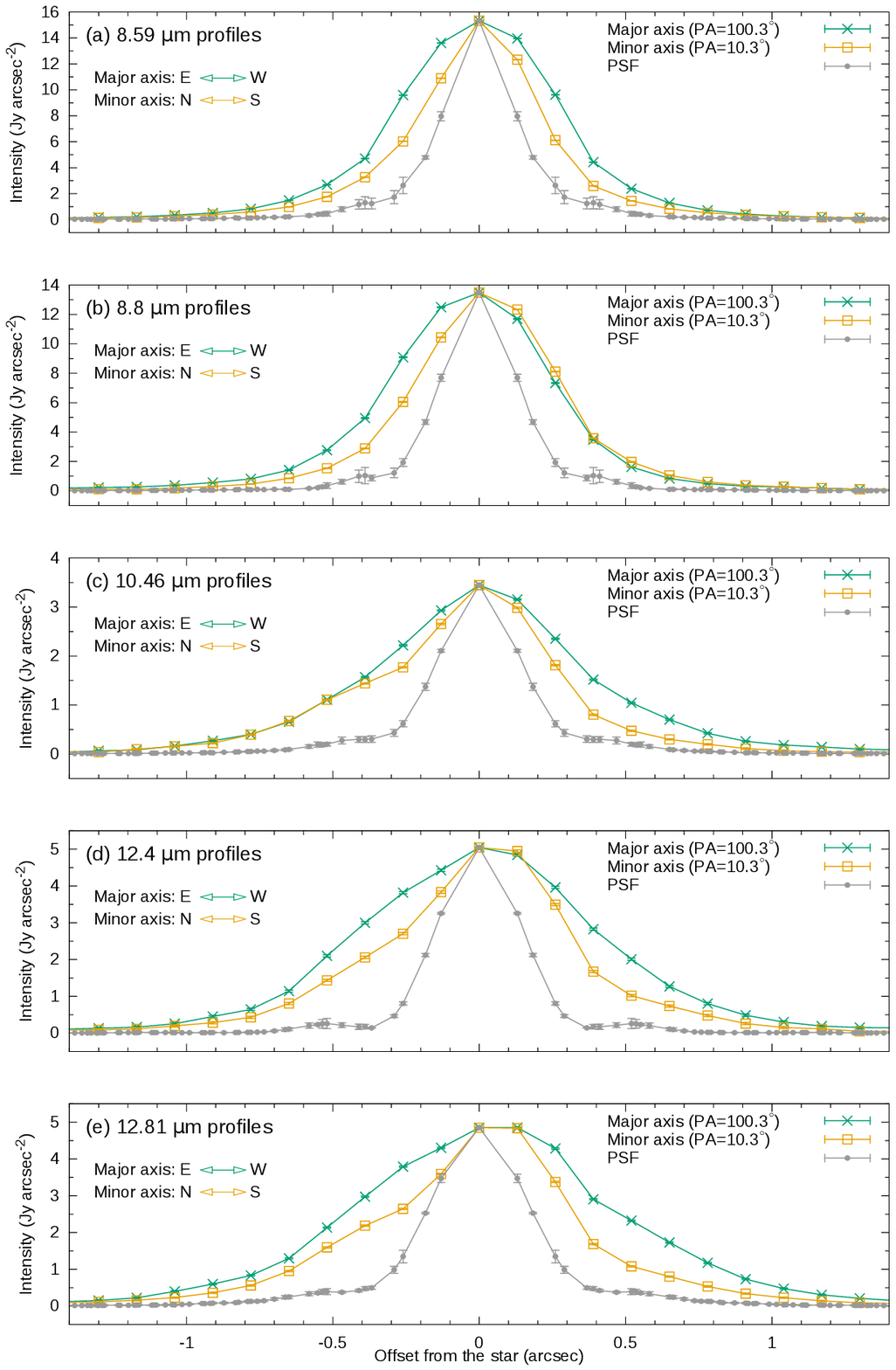}
\end{center}
 \caption{{\it N}-band radial brightness profiles of Oph\,IRS\,48. The green cross marks represent 
intensity at each angular distance from the center along the major axis, while the yellow squares 
along the minor axis. The grey circles show the point spread function scaled to the peak of IRS\,48. 
Note that the position alignment has an uncertainty of about 0.1 arcsec along each axis at every 
wavelength. The east- and west-side profiles are basically similar to each other. Their slight 
difference can be mostly explained by the position alignment error.}
\label{RadPro}
\end{figure*}

Figure \ref{RadPro} presents profiles of Oph\,IRS\,48 along the major or minor axis. 
Here we defined the direction of disk position angle (PA) of 100.3 degree from north to east according to the previous literatures 
\citep{gee07a,bru14a} and its perpendicular direction as the minor axis. 
The profiles at 8.59 and 8.8\,$\mu$m are nearly symmetric about the major and the minor 
axes, suggesting that those emissions come from a compact elliptical structure in the central 
region. On the other hand, the profiles at 10.46, 12.4, and 12.81\,$\mu$m are clearly asymmetric 
about the minor axes, which is consistent with the upside-down triangle shapes as seen in 
Figure \ref{IRS48images}. Furthermore, these profiles are relatively extended in comparison with those of the 
8.59 and 8.8\,$\mu$m, even after taking into consideration of relatively larger PSF sizes. 

The difference between the profiles at the three long wavelengths and those at the other two short 
wavelengths may be caused by the difference in the emitting dust species. A MIR spectrum 
of Oph\,IRS\,48 was obtained by Geers et al. \citep{gee07a,gee07b}. The spectrum shows PAH 
emissions at 7.7, 8.6, 11.3, and 12.7\,$\mu$m. Especially the 7.7\,$\mu$m feature 
is very prominent and it dominates the fluxes within the 8.59 and 8.8\,$\mu$m filters. On the other hand, 
the 11.3 and 12.7\,$\mu$m features are not adequately strong relative to the underlying continuum. 
Therefore, it is suggested that the PAHs are mainly distributed in the central region of the object, 
resulting in the compact shape of the images at 8.59 and 8.8\,$\mu$m.

The central shapes of the images at 12.4 and 12.81\,$\mu$m are slightly elongated in the east-west direction when compared to the image at 10.46\,$\mu$m. 
\citet{maa13a} constructed a model to represent a spectral energy distribution of 
Oph\,IRS\,48, and suggested that fluxes at the longer side of the {\it N}-band wavelengths 
have a contribution from the cooler dust in the outer disk component. 
Its contribution becomes larger as the wavelength is longer. This may naturally explain 
the spatial extents of the images at 12.4 and 12.81\,$\mu$m.

\subsection{{\it Q}-band Imaging}
\begin{figure*}
\begin{center}
 \includegraphics[width=160mm]{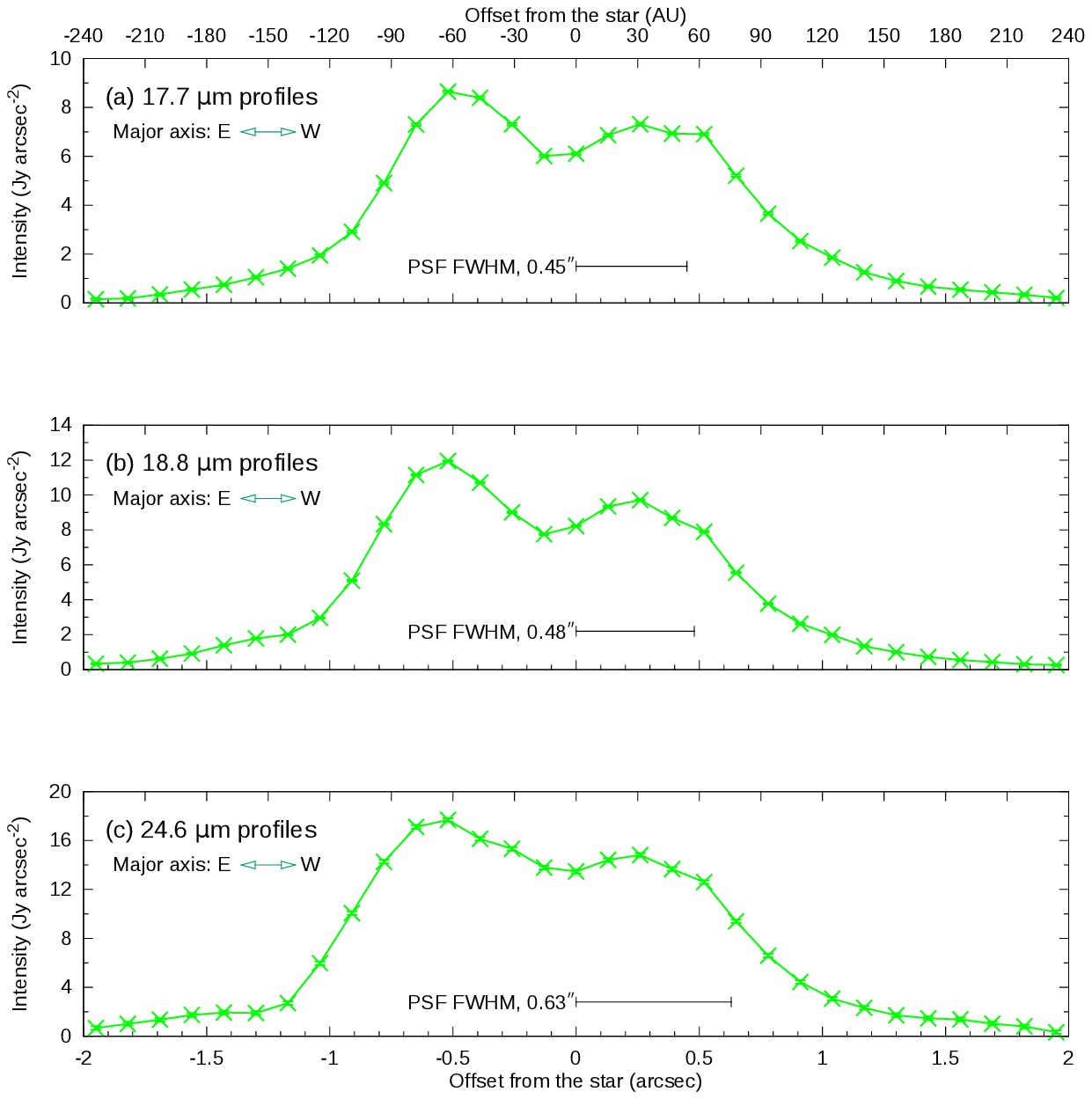}
\end{center}
 \caption{{\it Q}-band radial brightness profiles of Oph\,IRS\,48 cut along the major axis.}
\label{QRadPro}
\end{figure*}

Figure \ref{IRS48images} shows that the 17.7, 18.8 and 24.6\,$\mu$m emissions have a ring-like structure with a cavity 
in the center and double peaks along the major axis with a separation of $\sim$100 AU. 
These appearances were also observed in the VISIR 18.7\,$\mu$m image by \citet{gee07a}.
The disk has asymmetric brightness distributions in both the north-south and east-west directions. 
Figure \ref{QRadPro} displays the {\it Q}-band radial profiles of IRS\,48 cut along the major axis (PA = 100.3 deg) 
passing through the center of the images in Figure \ref{IRS48images}. 
The radial profiles are much alike, with slight differences. The western side has a slightly longer tail. 
The 24.6\,$\mu$m radial profile is smoother than the others although it is reasonable 
considering the broader PSF size at 24.6\,$\mu$m.
The brightness of the western peak is 81, 77, and 82 percent of the eastern flux peak at 17.7, 
18.8, and 24.6\,$\mu$m, respectively. The northern side of the disk, which is 50 deg inclined 
\citep{gee07a,bru14a}, is farther to us and more luminous than the 
southern side at every wavelength.


\section{Discussion}
\subsection{Origin of East-West Asymmetry}

\begin{figure*}
\begin{center}
 \includegraphics[width=70mm]{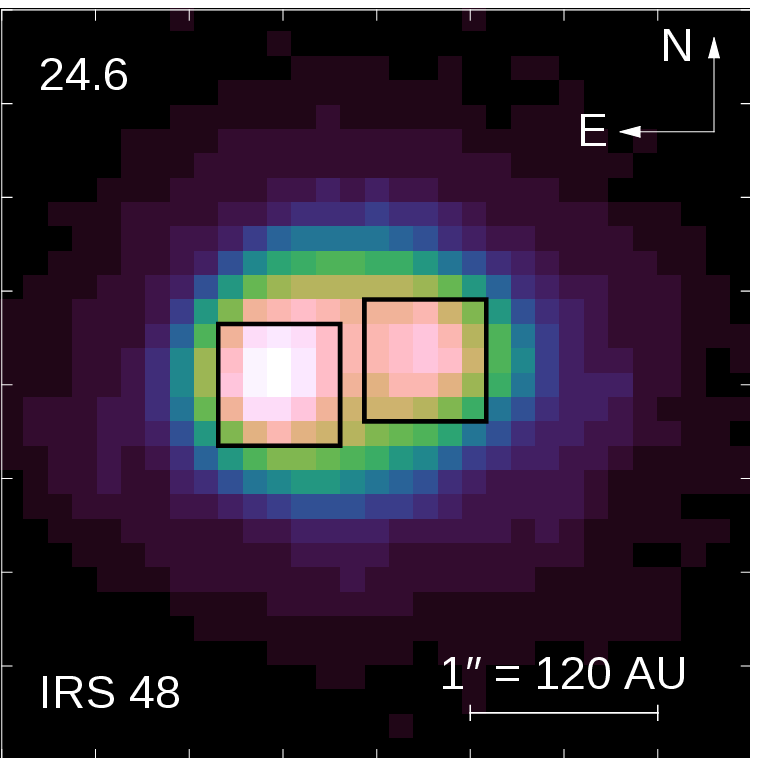}
\end{center}
 \caption{The square apertures centered at the east and west peaks overlaid in the 24.6\,$\mu$m image shown in Figure\ref{IRS48images}. The size of the square is 0.65''}
\label{EWcomp}
\end{figure*}

\cite{fol15a} pointed out that there might be an offset of disk cavity center from the stellar position, {\bf although their observations cannot confirm it.}
The observed east-west asymmetry in the {\it Q}-band could originate from the different temperature of the emitting dust. If this is the case, it implies that the inner wall radius is different between east and west, implying that the cavity is eccentric.
Since we have spatially resolved multiple {\it Q}-band images, we can derive the color temperature of the emitting dust at the east and west peak regions separately for the first time, and check if this hypothesis is true.
Unfortunately, due to the limited spatial resolution in the {\it Q}-band, we measured the averaged surface brightness in 18.8 and 24.6\,$\mu$m only in the square apertures centered at east and west peaks shown in Figure \ref{EWcomp}. The size of the square is 0.65'' which is comparable to the spatial resolution in 24.6\,$\mu$m.
To match the beam size between two images, gaussian beam convolution is applied to the 18.8\,$\mu$m image so as to match the FWHM of PSF to that of 24.6\,$\mu$m image. The measured averaged surface brightnesses for the east peak region are 9.3$\pm$1.4 Jy/arcsec$^2$ and 16.1$\pm$2.3 Jy/arcsec$^2$ at 18.8\,$\mu$m and 24.6\,$\mu$m, respectively,
while those for the west peak region are 7.8$\pm$1.2 Jy/arcsec$^2$ and 13.5$\pm$1.9 Jy/arcsec$^2$ at 18.8\,$\mu$m and 24.6\,$\mu$m, respectively. 
Coincidentally, the derived ratios ($I_{18.8}/I_{24.6}$) are 0.58$\pm$0.12 for both regions. 
This indicates that the temperature of the emitting dust at the east and west outer disk inner walls are not so different, and that the observed east-west asymmetry in the {\it Q}-band image may not be attributed to temperature difference. Thus, our results do not support the eccentric cavity scenario and are rather consistent with the symmetric cavity model.

Assuming the blackbody-like dust, the derived $I_{18.8}/I_{24.6}$ value corresponds to 135$\pm$22\,K, which is roughly consistent with the expected surface temperature of the inner wall of the outer disk \citep[Fig.5]{maa13a}. The origin of the east-west asymmetry in the {\it Q}-band is still unclear. It could be attributed to the different dust properties, different optical depth and/or small scale difference of structures such as scale height of the disk \citep{bru14a}. Furthermore, slight temperature difference ($\Delta T < 22K$) can not be ruled out yet, since our observations suffer from lower spatial resolution (0.65''). 
Future high spatial resolution Q-band observations ($<$0.1'') by ELTs are strongly desired for further study.

\subsection{Geometry of Oph\,IRS\,48}

\begin{figure*}
\begin{center}
 \includegraphics[width=160mm]{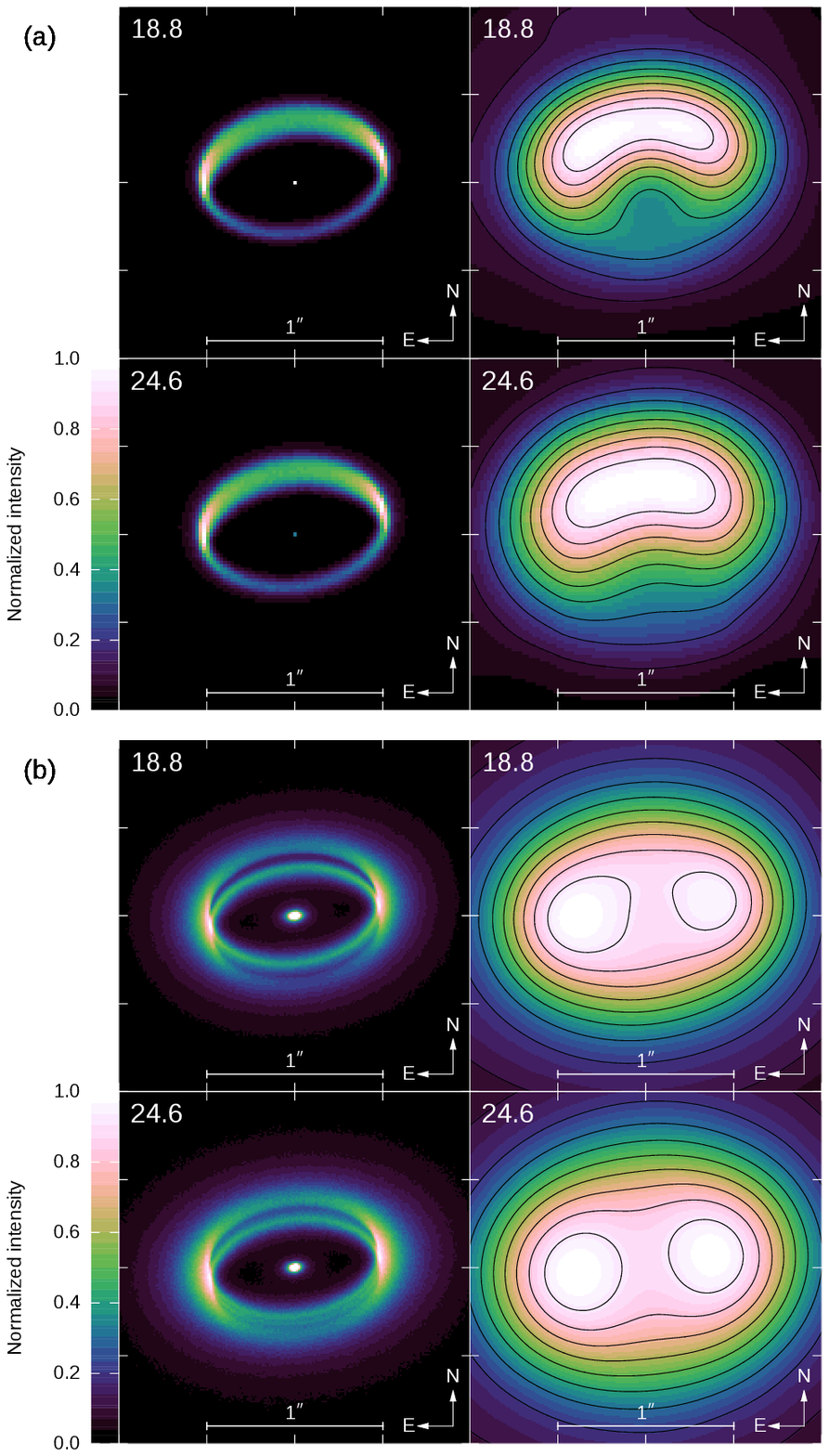}
\end{center}
 \caption{{\it Q}-band surface brightness distributions of Oph IRS 48 reconstructed from the models 
by (a) \citet{maa13a} and (b) \citet{fol15a}. Left: simulated images. Right: images 
convolved with the COMICS PSFs. Intensity in each panel is normalized to the peak of the outer disk component. Contours are overlaid at the levels of 10, 20, 30, 40, 50, 60, 70, 80, 90\,\% of the peak}
\label{models}
\end{figure*}

Radiative transfer modeling to reproduce the spatial distribution of a disk as well as SED is a powerful 
tool for investigating the disk structure. A number of studies have already been conducted for understanding 
the disk structure of IRS\,48 so far. Hereafter we compared several models with our observations. 
\citet{maa13a} constructed a model to reproduce its SED and the MIR emitting region size in
18.7\,$\mu$m taken by the VLT/VISIR, and found that a disk model consisting of an outer disk and an inner halo can explain them reasonably well. In their study, inner halo is needed to 
reproduce NIR flux that shows clear excess over the stellar photosphere. 
The left panels of Figure \ref{models}a displays the images at 18.8 and 24.6\,$\mu {\rm m}$ reconstructed by their model, and the right panels of Figure \ref{models}a are the images convolved with the PSF of the Subaru/COMICS.
The horseshoe-like structure seen in the PSF-convolved images is inconsistent with the 
observed double peaked morphology along the semi-major axis of the ring.

Recently \citet{fol15a} conducted AO observations at {\it H}- and {\it K}$_{\rm s}$-bands and presented polarized intensity (PI)
images that consist of two pronounced arcs with radii of $\sim 60$ AU. They also built a radiative transfer model that 
accounts for the NIR images and the SED. In their model, a structure with inner and outer disks was adopted to explain the arc structures. Figure \ref{models}b shows images simulated by this model at 18.8 and 24.6\,$\mu {\rm m}$ and the PSF convolved images. This model successfully reproduces the double peaked morphology along the semi-major axis of the ring in {\it Q}-band. {\bf Note that the east-west asymmetry slightly seen in Figure \ref{models}b 18.8\,$\mu$m PSF convolved image is due to the slight asymmetry of the COMICS 18.8\,$\mu$m PSF displayed in Figure \ref{IRS48images}. The original model image is symmetric about the disk minor axis.} There are several differences between disk models of \citet{maa13a} and those of \citet{fol15a}. The main difference is the assumption of the inner structure.
\citet{maa13a} assumed an inner optically thin halo, while \citet{fol15a} used an optically thick inner disk. If the inner components were optically thick, they would cast a shadow on the wall of the outer disk. This strongly affects the physical structure of the disk such as temperature and scale height, 
and thus, influences the infrared morphology of the system (see more detailed discussions in \citet{mulders2010}).

By comparing two previous modeling works with different assumptions on the inner disk structures, the optically thick inner disk model seems to be the better solution for the IRS\, 48 case.
The inner disk should cast a shadow over the outer disk wall and 
the surface brightness is suppressed in the disk far-side wall, which is directly exposed to us, 
resulting in the double peak morphology along the major axis. 
A similar modeling result is obtained by \citet{bru14a} for this system, supporting above interpretation for this system.

\subsection{Comparison with HD\,142527 mid-IR morphology interpretation}
We would like to point out the difference of interpretation of MIR morphology between Oph\,IRS\,48 case and 
another well-studied Herbig Fe star, HD\,142527. HD\.142527 has a transitional disk with an inclination of 28 degrees \citep{per14a}. 
The disk possesses an asymmetric dust depleted gap ($\sim 140$ AU radius) and spiral arms \citep{fuk06a,fuj06a,ave14a}. 
ALMA observations of dust continuum show asymmetric horseshoe-like emission in the northern part of the disk, 
which can be interpreted as a result of cavity clearing by giant planets or a dust trapping 
at the local gas pressure maxima \citep{cas13a,fuk13a}. In the MIR wavelengths ({\it Q}-band), HD\,142527 shows a strong central source (inner disk) with east and west arcs (outer disk) around it \citep{fuj06a,ver11a}. Eastern arc is brighter than western arc, and east-west direction is almost the outer disk minor axis (major axis position angle PA=160$^\circ$).
To interpret this brightness asymmetry of the east and west arcs, \citet{fuj06a} explained
that the eastern side is the far side from us
and the inner rim/wall of the outer disk on the eastern side is directly exposed to us, 
while western side suffers from extinction by outer disk.
This situation resembles our PSF-convolved IRS 48 model images of \citet{maa13a} in right panels of Figure \ref{models}a, showing that the far side of the outer disk is the brightest.
Thus, the interpretations of the disk morphology are different between Oph\,IRS\,48 and HD\,142527.

The main difference between Oph\,IRS\,48 and HD\,142527 seems to be the (mis)alignment of inner and outer disks. Recently, \citet{mar15a} suggested that the inner disk of HD142527 is misaligned to the outer disk to explain two intensity nulls at outer disk at position angles of 0 and 160 degree \citep{ave14a}. 
The optically thick misaligned disk casts deep shadows on the two positions of the outer disks, resulting in the two arc structures seen in the NIR PI image. 
This misalignment also accounts for the full illumination of the far side wall of outer disk of HD\,142527, which result in brighter far side wall in the infrared wavelengths.
The misaligned disk scenario for HD\,142527 is also reported by ALMA observations \citep{cas15a}.
On the other hand, Oph\,IRS\,48 inner disk is supposed to be aligned to the outer disk.
The optically thick inner disk casts its shadow at the outer disk wall as seen in Fig. \ref{models}b left panels, resulting in suppression of the far side wall brightness.
This is a simple and thus preferable explanation for the difference of the interpretations of the disk morphology. 


To explain the observed disk structures, presence of companions/planets are suggested to both sources. Meanwhile, HD\,142527 has a stellar companion whose mass is estimated 0.1 to 0.4 $M_\odot$ \citep{bil12a}. \citet{lac16a} 
analyze physical parameters including orbital elements and eventually indicate that the direction of the companion's orbit is the 
same as the tilt of the inner disk. This is a strong support for the dynamical interaction scenario. On the other hand, it is not clear 
whether Oph\,IRS\,48 has companions. \citet{mar13a} suggested that a massive planet or brown dwarf ($\sim 10$ $M_{\rm Jupiter}$) 
caused a dust trap seen as in the highly azimuthally asymmetric distribution in the sub-millimeter continuum. Presence of companions are also suggested by other studies \citep{bro12a,bru14a,mar16}.
However, a companion is not directly observed so far for Oph\,IRS\,48. 
Further high resolution observations are needed to 
clarify the origin of the inner structure of the Oph\,IRS\,48 disk. This also contributes further understanding of the disk structure and planet formation process in the disk.

\begin{ack}
We are grateful to all the staff members of the Subaru Telescope.
We also thank the referee, Dr. Nienke van der Marel, for her useful comments.
MH was supported by JSPS/MEXT KAKENHI Grant Number 15H05439.
This study was supported by the Grant of Joint Development Research supported by the Research Coordination Committee in National Astronomical Observatory of Japan (NAOJ), the Grant for research and development of TMT instruments from NAOJ, and the grant from the Ishibashi Foundation for the Promotion of Science, 

\end{ack}




\end{document}